# Flexible Tuning of Asymmetric Near-field Radiative Thermal Transistor by Utilizing Distinct Phase Change Materials

Hexiang Zhang[1], Xuguang Zhang[1], Fangqi Chen[2], Mauro Antezza[3,4], Yi Zheng[1,5,*]

[1]*Department of Mechanical and Industrial Engineering, Northeastern University, Boston, MA 02115, USA*

[2]*Department of Mechanical Engineering, University of Michigan, Ann Arbor, MI 48109, USA*

[3]*Laboratoire Charles Coulomb (L2C), UMR 5221 CNRS-Université de Montpellier, F-34095 Montpellier, France*

[4]*Institut Universitaire de France, 1 rue Descartes, Paris Cedex 05 F-75231, France*

[5]*Department of Chemical Engineering, Northeastern University, Boston, MA 02115, USA*

[*]**Author to whom correspondence should be addressed:** y.zheng@northeastern.edu

## ABSTRACT

Phase change materials (PCMs) play a pivotal role in the development of advanced thermal devices due to their reversible phase transitions, which drastically modify their thermal and optical properties. In this study, we present an effective dynamic thermal transistor with an asymmetric design that employs distinct PCMs, vanadium dioxide ($VO_2$) and germanium antimony telluride (GST), on either side of the gate terminal, which is the center of the control unit of the near-field thermal transistor. This asymmetry introduces unique thermal modulation capabilities, taking control of thermal radiation in the near-field regime. $VO_2$ transitions from an insulating to a metallic state, while GST undergoes a reversible switch between amorphous (aGST) and crystalline (cGST) phases, each inducing substantial changes in thermal transport properties. By strategically combining these materials, the transistor exhibits enhanced functionality, dynamically switching between states of absorbing and releasing heat by tuning the temperature of gate. This gate terminal not only enables active and efficient thermal management but also provides effective opportunities for manipulating heat flow in radiative thermal circuits. Our findings highlight the potential of such asymmetrically structured thermal transistors in advancing applications across microelectronics, high-speed data processing, and sustainable energy systems, where precise and responsive thermal control is critical for performance and efficiency.

In recent years, the development of near-field radiative heat transfer (NFRHT) technology has provided strong support for breaking through the bottleneck of traditional thermal management methods.[1,2] Near-field radiation can achieve radiant heat transfer above the blackbody limit at very short distances, to achieve efficient heat exchange and regulation at the nanoscale.[3] The core of this kind of technology lies in the precise tuning of the radiation characteristics of the material, and PCM has gradually become the ideal material choice for the realization of micro-nano thermal control devices because of its unique phase transition ability and repeatable thermo-optical properties.[4,5,6]

Near-field thermal diodes and thermal transistors are two key components in thermal management systems[7], and their working principle is similar to that of electrical diodes and transistors in electronic devices.[8] In the process of heat transmission, the energy is preferentially transmitted in a single direction, to achieve the rectification function of heat flow.[9] This property is particularly important in the thermal circuit, helping to achieve directional control of heat flow. Thermal diodes are usually achieved through specific structural design or the temperature-dependent thermal properties of the material.[10,11,12] The thermal transistor, as a more advanced thermal regulatory element, can realize the dynamic regulation of heat flow according to external control signals (such as temperature, voltage or mechanical deformation).[13] Thermal transistors can quickly switch between high and low thermal conductivity states, to achieve the current regulation function similar to electronic transistors, providing an effective solution for active thermal management.

The key to realize dynamically adjustable near-field radiant thermal management devices lies in the application of PCM. VO$_2$ and GST alloys, as two typical PCMs, stand out for their unique phase transition properties.[14] During the phase transition, VO$_2$ can transition from an insulating state to a metallic state at about 341K, with dramatic changes in electrical and thermal conductivity[15]; GST can achieve reversible transition between crystalline and amorphous states, changing its optical constant and heat conduction properties.[16,17,18] By using the controllable phase transformation characteristics of these PCM, the dynamic control of near-field radiant heat flow can be realized, thus promoting the development of active thermal management devices such as thermal transistors.[19]

In this research, we propose a tunable near-field thermal radiation transistor design based on asymmetric nanostructures. The structure is composed of VO$_2$ and GST diffraction gratings integrated on both sides, and realizes real-time control of near-field radiation heat flow through the unique characteristics of phase change materials.[20] Through external thermal or optical stimulation, VO$_2$ and GST can switch between different phase states, so that the transistor gate can dynamically adjust the heat flow, and realize the flexible switch of heat absorption and heat release functions.[21] This design not only provides a fundamental method of active thermal management, but also makes up for the limitations of passive heat dissipation technology and provides higher precision and flexibility for nanoscale heat flow regulation.[22]

Before showing the transistor structure and results, theoretical fundamentals are necessary to be recalled.[23,24] Dyadic Green's function is employed to calculate the radiative transfer between two near-field planar surfaces for the thermal transistor[25]:

$$Q_{1\to2} = \int_0^\infty \frac{d\omega}{2\pi} [\Theta(\omega, T_1) - \Theta(\omega, T_2)] T_{1\to2}(\omega, L) \tag{1}$$

The energy of harmonic oscillator at frequency $\omega$ and temperature $T$:

$$\Theta(\omega, T) = \frac{\hbar\omega}{2} \coth\left(\frac{\hbar\omega}{2k_B T}\right) \tag{2}$$

where $\hbar$ is reduced Planck constant, and $k_B$ is the Boltzmann constant.

The spectral radiative transmissivity between two media by distance $L$:

$$T_{1\to2}(\omega, L) = \int_0^\infty \frac{k_\rho dk_\rho}{2\pi} \xi(\omega, k_\rho) \tag{3}$$

where $k_\rho$ is the parallel component of wavevector.

The energy transmission coefficient is defined as:

$$\xi\left(\omega, k_\rho \leq \frac{\omega}{c}\right) = \sum_{\mu=s,p} \frac{(1-\left|\tilde{R}_1^{(\mu)}\right|^2)(1-\left|\tilde{R}_2^{(\mu)}\right|^2)}{\left|1-\tilde{R}_1^{(\mu)}\tilde{R}_2^{(\mu)}e^{2jk_zL}\right|^2} \tag{4}$$

$$\xi\left(\omega, k_\rho > \frac{\omega}{c}\right) = \sum_{\mu=s,p} \frac{4\Im(\tilde{R}_1^{(\mu)})\Im(\tilde{R}_2^{(\mu)})e^{-2|k_z|L}}{\left|1-\tilde{R}_1^{(\mu)}\tilde{R}_2^{(\mu)}e^{-2|k_z|L}\right|^2} \tag{5}$$

where $\tilde{R}_1^{(\mu)}$ and $\tilde{R}_2^{(\mu)}$ are two half spaces, polarization dependent reflection coefficients. $\mu = s$ (or $p$) is the transverse electric (or magnetic) polarization. $k_z$ is the $z$-component of wavevector in vacuum. Propagating and evanescent modes are corresponding to $k_\rho \leq \frac{\omega}{c}$ and $k_\rho > \frac{\omega}{c}$, respectively.

For 1-D grating PCM structure in vacuum, the second order approximation of effective medium theory is used to generate the related dielectric properties:

$$\varepsilon_{TE,2} = \varepsilon_{TE,0}\left[1 + \frac{\pi^2}{3}\left(\frac{\Lambda}{\lambda}\right)^2 \phi^2(1-\phi)^2\frac{(\varepsilon_A-\varepsilon_B)^2}{\varepsilon_{TE,0}}\right] \tag{6}$$

$$\varepsilon_{TM,2} = \varepsilon_{TM,0}\left[1 + \frac{\pi^2}{3}\left(\frac{\Lambda}{\lambda}\right)^2 \phi^2(1-\phi)^2(\varepsilon_A-\varepsilon_B)^2\varepsilon_{TM,0}\left(\frac{\varepsilon_{TM,0}}{\varepsilon_A\varepsilon_B}\right)^2\right] \tag{7}$$

$\varepsilon_A$ and $\varepsilon_B$ are dielectric functions of the two materials (PCM and vacuum) in surface grating, $\lambda$ is the wavelength, $\Lambda$ is grating period, $w$ is the width of PCM segment; Therefore, the filling ratio can be defined as $\phi = \frac{w}{\Lambda}$.

The functions for zeroth order effective dielectric functions:

$$\varepsilon_{TE,0} = \phi\varepsilon_A + (1-\phi)\varepsilon_B \tag{8}$$

$$\varepsilon_{TM,0} = \left(\frac{\phi}{\varepsilon_A} + \frac{1-\phi}{\varepsilon_B}\right)^{-1} \tag{9}$$

The phase change process isn't linear. The Arrhenius equation, Sigmoid function and tanh function are suitable to simulate the phase-change process, here tanh function is chosen.[26,27]

$$f_A(T) = \frac{1}{2}\left[1 - \tanh\left(\frac{T-T_c}{k_m \cdot \Delta T}\right)\right] \tag{10}$$

$$f_B(T) = \frac{1}{2}\left[1 - \tanh\left(\frac{T-T_c}{k_m \cdot \Delta T}\right)\right] \tag{11}$$

Here, $f_A(T)$ and $f_B(T)$ are the percentage of volume of low and high temperature phase, respectively. $T_c$ is the stagnation temperature of phase change. $\Delta T$ is the total length of temperature for phase change. $k_m$ is the modified factor to reasonably describe the phase-change process. Based on the multiple tests, $k_m$ for VO$_2$ and GST will be chosen as 0.1. In this case, we would choose $\Delta T_{VO_2} = 5K$ from anisotropic insulator state to metallic state begin at 341K, $\Delta T_{GST} = 20K$ from amorphous-phase to crystalline-phase begin at 432K, and don't consider the hysteresis effect because of one direction phase change. Then, Maxwell Garnett effective medium theory is used to get the equivalent dielectric constant during the phase-change process:

$$\frac{\varepsilon_{\text{eff}} - \varepsilon_m}{\varepsilon_{\text{eff}} + 2\varepsilon_m} = \delta_i \frac{\varepsilon_i - \varepsilon_m}{\varepsilon_i + 2\varepsilon_m} \tag{12}$$

$\delta_i$ is the volume fraction of inclusions. $\varepsilon_m$ and $\varepsilon_i$ are the dielectric constants of the matrix and inclusions, respectively. $\varepsilon_{\text{eff}}$ is the effective dielectric constant of the medium. Based on the parameter of the structure shown on Fig. 1(b), $\delta_i < 10^{-5}$; Therefore, the Maxwell Garnett formula is correct.

The thermal transistor can be seen as two thermal diodes combined in an inverse direction. For the initial near-field structure without the gate, the length of the gap is $L = 100$nm. Fig. 1(a) shows the near-field structure of the thermal transistor with three regions source, gate, and drain from left to right. BN (blue) and Au (yellow) are in the regions of source and drain. VO$_2$ (red) and GST (brown) are covered on the gate region. The color of dark blue is representative of substrate. $d$ is the length of the gap distance from the gate to the opposite sides, which is half of the initial length of $L$. The thickness for all grating and thin film structure are $t_1 = 0.5$μm and $t_2 = 1$μm, respectively. The grating period (Λ) of the 1-D grating structure is 50nm with filling ratio ($\phi$) equal to 0.3. Fig. 1(b) presents three additional asymmetric gate design schemes utilizing PCMs, highlighting four distinct asymmetric gate configurations.

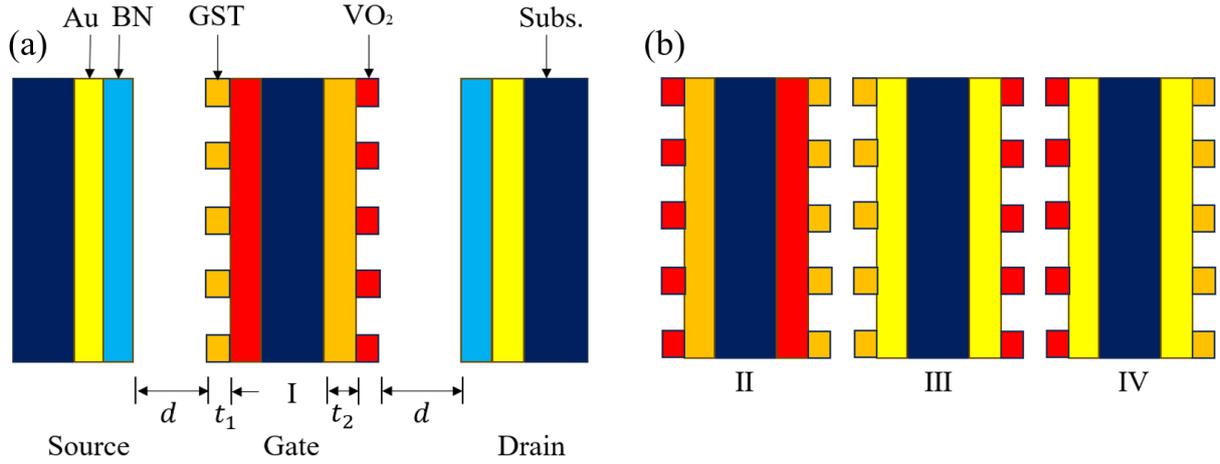

FIG. 1. (a) The schematic structure of near-field thermal transistor. (b) Three other different asymmetrical gates. The materials used: VO2 (red), GST (brown), BN (blue), Au (yellow) and Substrate (dark blue).

Figure 2 illustrates the values of $n$ and $\kappa$ of different states of GST and VO$_2$. Related data about VO$_2$ can be calculated by the dielectric function[5], and gotten from Frantz, Jesse A. et al[28] about GST.

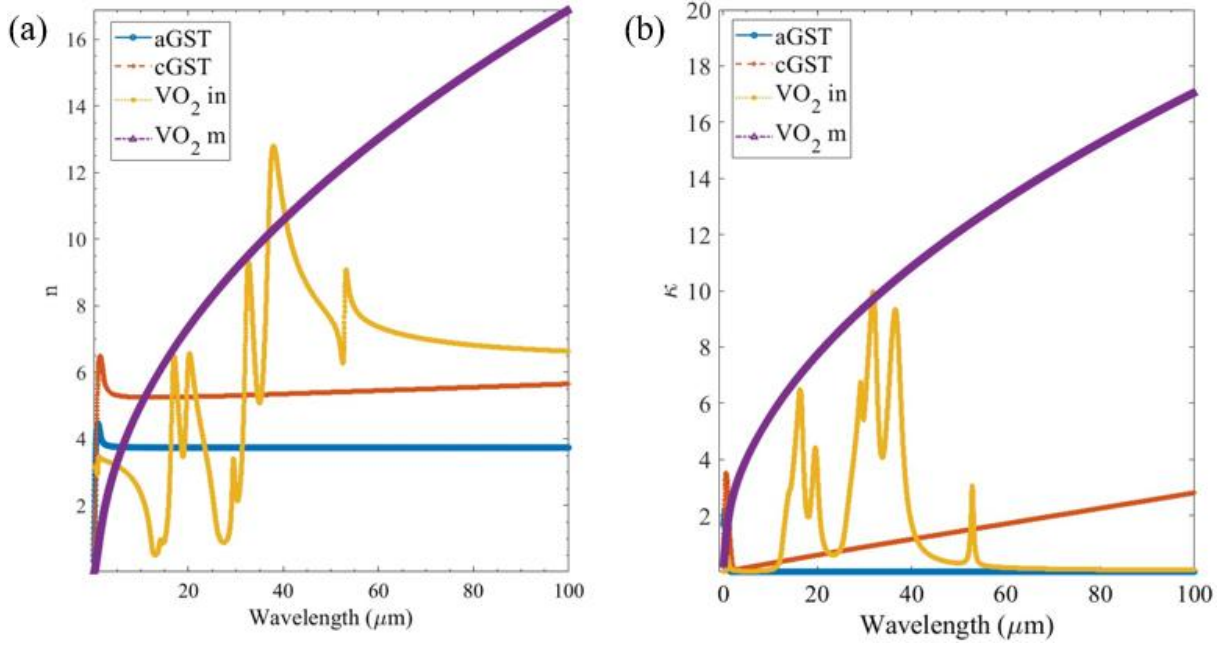

FIG. 2. Values of $n$ and $\kappa$ about different states of GST and VO₂. (a) Values of $n$. (b) Values of $\kappa$. VO₂ in and VO₂ m in the figure represent insulating state and metal state, respectively.

Figure 3 is the energy transmission coefficient $\xi(\omega, k_\rho)$ contour plot for crossing 1-D rectangular thermal diode with two interfaces. To analyze the operating mechanism of thermal diode can help us understand the extension of the thermal device, thermal transistor. The energy transmission coefficient of VO₂ grating with Au can be found in Ghanekar et al[5]. $k_\rho c/\omega$ is the normalized parallel wavevector. The energy transmission coefficient across the interfaces at a gap of 100nm is governed by the interplay of surface phonon polaritons and material properties. Due to aGST's ability to support surface phonon modes, dominates energy transmission in specific frequency ranges corresponding to its characteristic phonon wavelengths. These modes facilitate efficient tunneling of surface waves, particularly when the opposite interface exhibits a high extinction coefficient at these frequencies. However, cGST has a significantly different optical response, where the mismatch in phonon frequencies between amorphous and cGST suppresses resonant tunneling. cGST has a different way to support surface phonon polaritons in the same infrared region as aGST, and its extinction coefficient profile often limits energy transmission. This mismatch leads to lower overall energy transmission when the interface involves cGST. Similarly, VO₂ also plays a critical role due to its metal-insulator phase transition. In its metallic phase, VO₂ exhibits high extinction coefficients and does not support surface phonon polaritons in the infrared region. As a result, the tunneling between all states of GST and metallic VO₂ is primarily driven by the surface modes of GST. The symmetric and asymmetric surface phonons of aGST dominate the transmission when coupled with metallic VO₂, as it allows strong tunneling across interfaces. In the insulating phase of VO₂, surface phonon modes are supported, but their spectral range often does not overlap with those of GST. For example, the surface phonons of aGST typically occur in a range where insulating VO₂ exhibits low extinction coefficients, limiting resonant energy transfer. This mismatch further suppresses tunneling between GST and insulating VO₂, resulting in reduced energy transmission. When a 1-D rectangular grating replaces bulk VO₂, additional suppression occurs, particularly in configurations involving insulating VO₂. The grating disrupts the propagation of frustrated modes and Fabry-Perot resonances, further reducing the tunneling efficiency. However, the tunneling between aGST and metallic VO₂ remains relatively unaffected, as the metallic VO₂ supports broader interaction modes.[29]

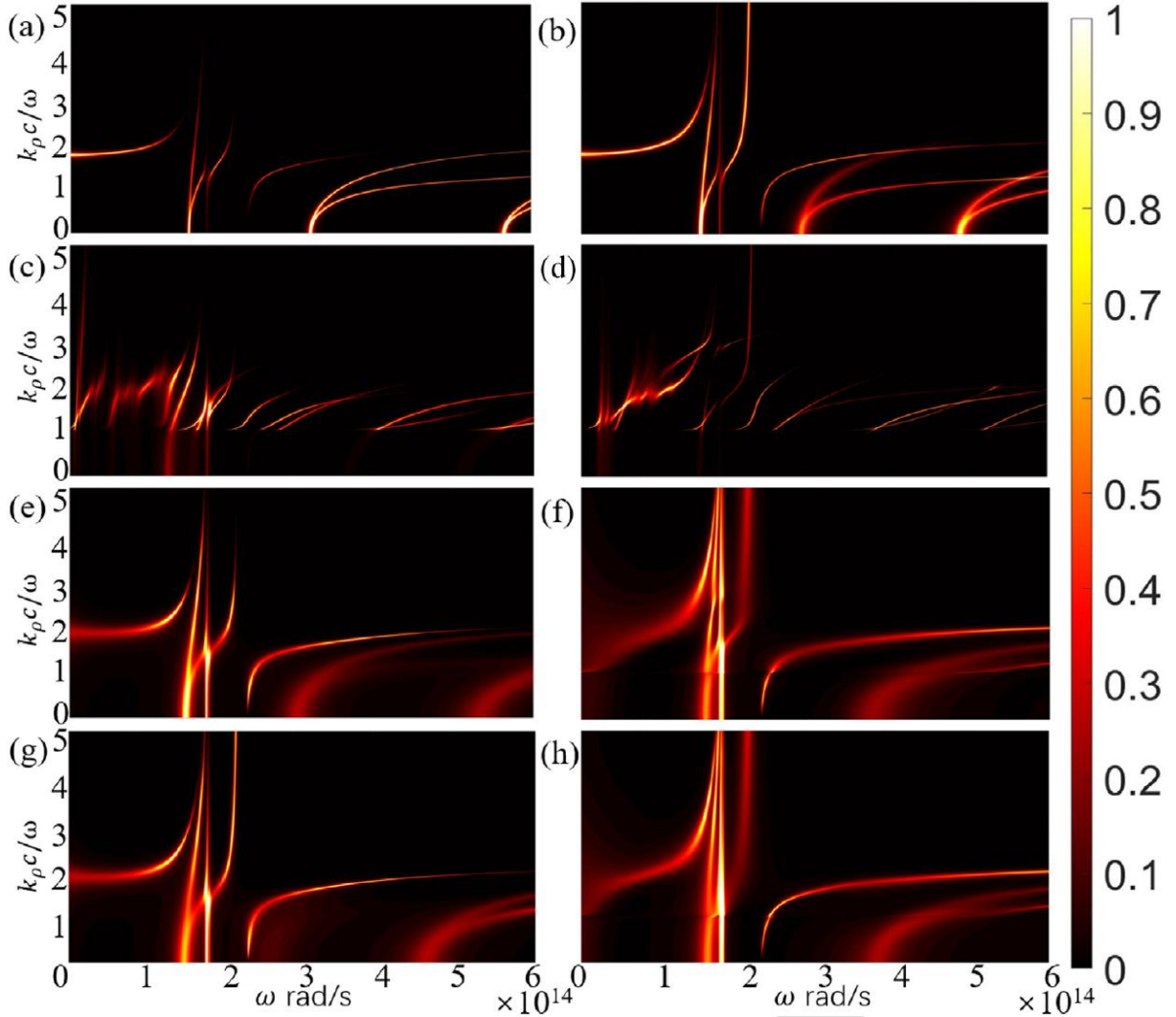

FIG. 3. Contour of energy transmission coefficient $\xi(\omega, k_\rho)$ across the two interfaces of thermal diode with 1-D rectangular grating structure. Each figure's unit is same, figure (g) for example, the vertical and horizontal coordinates represent the $k_\rho c/\omega$ and frequency, respectively. And the right rule is to represent the energy transmission coefficient. (a) aGST and (b) cGST grating covered on VO₂ thin film. aGST grating covered on (c) VO₂ insulating state and (e) metal state thin film. VO₂ (d) insulating state and (f) metal state covered on aGST thin film. (g) cGST grating covered on VO₂ metal thin film. (h) VO₂ metal state grating covered on cGST thin film.

Based on Fig. 1(a), we calculate the initial heat flux transfer from the source to the drain directly without gate added is $Q_{\text{initial}} = 836.854 \text{W/m2}$. After adding the gate between the source and drain, the transferring heat flux to the drain can be increased to almost 10 more times by tuning the temperature of the gate. Fig. 4 is to describe the specific change of the amplification factor. We will define the amplification factor as $\alpha = \frac{Q_D}{Q_{\text{initial}}}$, where $Q_G$ is the energy transfer from the gate to the drain in the thermal transistor.

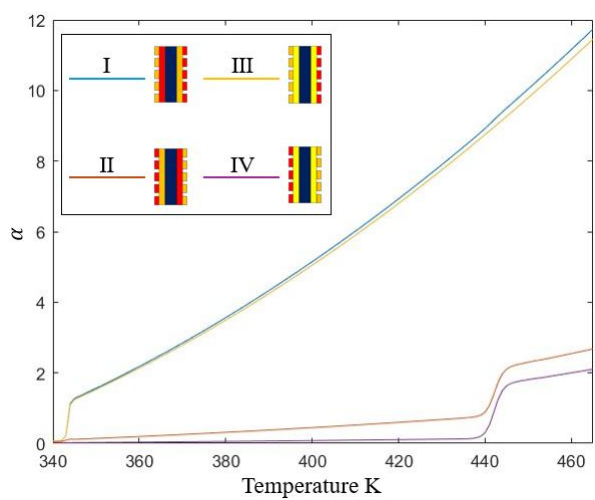

FIG.4 Tuning the gate temperature from 340K to 465K, the amplification factors gradually increase. The amplification factor of VO$_2$ grating surface on the gate (I, III) toward the drain increase markedly than the GST grating surface (II, IV). The grating structure covered on film hardly affects the amplification factor.

Figure 5 illustrates the heat flux across the source ($Q_S$), the gate ($Q_G$), and the drain ($Q_D$) as the gate temperature is tuned for four different structures. The relationship $Q_G = Q_S - Q_D$ is maintained due to the steady-state operation of the system. Analysis of the plots and variations in the amplification factor ($\alpha$) for the four asymmetric near-field thermal transistors reveals that the grating surface on the gate facing the drain plays a pivotal role in determining the amplification factor, significantly influencing heat transfer behaviors within the thermal transistor, as shown in Fig. 4. Notably, when $Q_G = 0$, a unique case emerges. In this condition, the source and gate regions effectively behave as a unified, enhanced amplification near-field thermal device. This phenomenon arises due to the thermal property changes in the PCMs[30], allowing the system to achieve a state where no heat flux passes through the gate. The absence of $Q_G$ transforms the thermal transistor's configuration, emphasizing the coupling between the source and drain regions and enabling highly efficient near-field heat transfer.[31]

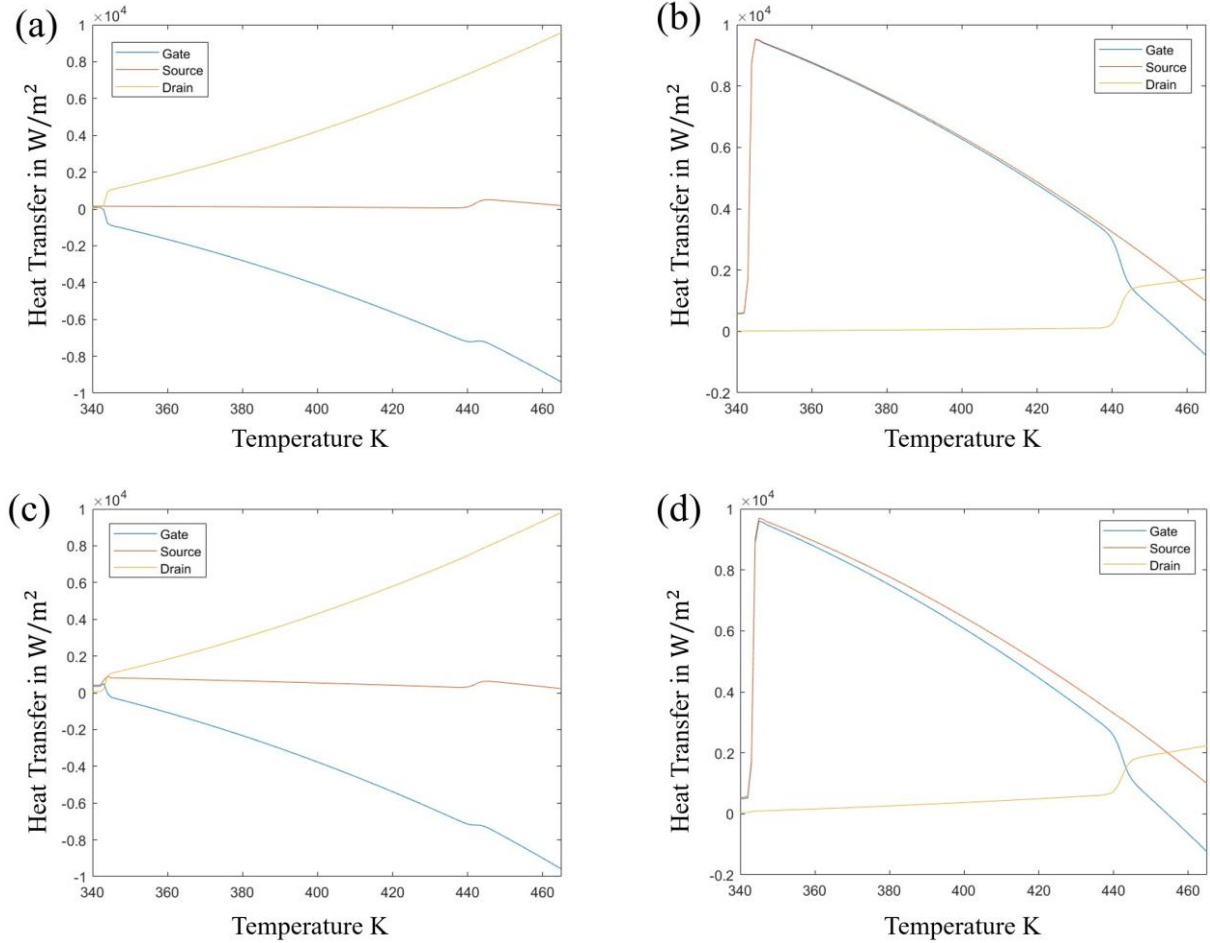

FIG.5. The heat transfer of different regions about structures (a) I. (b) II. (c) III. (d) IV, which has been shown in Fig.1.

In summary, based on the results derived from employing different PCMs to construct various surfaces on the gate, it is evident that when tuning the gate temperature results in a significant enhancement of the amplification factors.[32] Notably, the amplification factor associated with the VO₂ grating surface on the gate towards the drain exhibits a significantly greater increase compared to that of the GST grating surface. In addition, the grating structure covered thin film appears to have a negligible influence on the amplification factor. Also, in the condition of $Q_G = 0$ (namely, $Q_S = Q_D$), the regions of source and gate can function as a unified and enhanced amplification near-field thermal device, which can be simplified a thermal diode. This phenomenon suggests that the combination of multiple phase change materials (PCMs) may significantly enhance the likelihood of achieving such a state. Different PCMs, with varying phase transition temperatures and distinct thermal and optical properties, could work synergistically to finely tune the thermal flux distribution. For example, a tailored arrangement of VO₂ and GST, or other PCMs with complementary characteristics, might further optimize the heat transfer pathways and enable advanced control of thermal amplification. In this case, it can tune the gate terminal to reach a state where there is no heat exchange between the gate and the external environment. Such a state could serve as a critical design parameter in thermal management, enabling the gate terminal to act as a passive thermal insulator or active modulator depending on the operational requirements. By strategically combining these materials, the transistor enhances its functionality by dynamically switching between heat absorption and release through gate temperature tuning, thereby opening another avenues for highly responsive thermal management systems.

Moreover, this approach could pave the way for the development of multifunctional thermal devices, where precise control over $Q_G$ enables applications such as thermal logic gates, thermal memory devices, and adaptive thermal shields in electronics and energy systems. And, we can also utilize the properties of the change of $Q_G$ to control the supplying or releasing heat. This finding provides additional options for the design and construction of thermal circuits with specific requirements. This work was financially supported by the National Science Foundation (Grant number: CBET-1941743).

## AUTHOR DECLARATIONS

### Conflict of Interest

### Author Contributions
**Hexiang Zhang:** Conceptualization (equal); Data curation (lead); Methodology (lead); Validation (lead); Writing – original draft (lead); Writing – review & editing (equal). **Xuguang Zhang:** Writing – review & editing (equal). **Fangqi Chen:** Data curation (supporting), Writing – review & editing (equal). **Mauro Antezza:** Writing – review & editing (equal). **Yi Zheng:** Supervision (lead); Project administration (lead); Funding acquisition (lead) Conceptualization (equal); Writing – review & editing (equal).

## DATA AVAILABILITY

The data that support the findings of this study are available from the corresponding authors upon reasonable request.